\documentclass[aps,prd,preprintnumbers,nofootinbib,twocolumn]{revtex4}
\usepackage{bm}
\usepackage{latexsym}
\usepackage{dcolumn}
\usepackage{amsmath,amsfonts,amssymb}
\usepackage{graphicx,epsfig}
\usepackage{psfrag}
\usepackage{amsthm}
\usepackage{color}

\usepackage{bm}
\usepackage{latexsym}
\usepackage{dcolumn}
\usepackage{amsmath,amsfonts,amssymb}
\usepackage{graphicx,epsfig}
\usepackage{psfrag}

\usepackage{nccmath}
\usepackage{moresize}
\usepackage{enumerate}
\usepackage[hang,flushmargin]{footmisc}
\usepackage[titletoc,toc]{appendix}
\usepackage{lipsum}
\usepackage{epstopdf} 
\usepackage{hyperref}
\usepackage{rotating}
\usepackage{multirow}
\usepackage{mathtools}

\usepackage{stackengine}
\usepackage{scalerel}

\newcommand{\be}{\begin{equation}}
\newcommand{\ee}{\end{equation}}
\newcommand{\bea}{\begin{eqnarray}}
\newcommand{\eea}{\end{eqnarray}}
\newcommand{\bse}{\begin{subequations}}
\newcommand{\ese}{\end{subequations}}
\newcommand{\bce}{\begin{center}}
\newcommand{\ece}{\end{center}}
\newcommand{\bfg}{\begin{figure}}
\newcommand{\efg}{\end{figure}}
\newcommand{\bit}{\begin{itemize}}
\newcommand{\eit}{\end{itemize}}
\newcommand{\bed}{\begin{description}}
\newcommand{\eed}{\end{description}}
\newcommand{\ben}{\begin{enumerate}}
\newcommand{\een}{\end{enumerate}}
\newcommand{\nn}{\nonumber}

\newcommand{\fr}{\frac}
\newcommand{\sq}{\sqrt}
\newcommand{\no}{\noindent}

\def\le {\left}
\def\ri {\right}
\def\d  {\delta}

\def\e  {\epsilon}

\def\l  {\lambda}
\def\L  {\Lambda}

\def\r  {\rho}

\newcommand{\cR}{\mathcal R}

\newcommand{\vx}{\vec{\pmb x}}

\newcommand{\nabtr}{\vec\nabla\!_{\perp}}

\newcommand{\bdm}{\begin{displaymath}}
\newcommand{\edm}{\end{displaymath}}

\begin{document}

\title{Gravitational lensing and missing mass}

\author{Saurya Das} \email[email: ]{saurya.das@uleth.ca}
%%
%

%\vs{0.3cm}

\affiliation{Theoretical Physics Group and Quantum Alberta,
Department of Physics and Astronomy,
University of Lethbridge, 4401 University Drive,
Lethbridge, Alberta T1K 3M4, Canada}

\author{Sourav Sur}
\email[email: ]{sourav.sur@gmail.com, sourav@physics.du.ac.in}

\affiliation{Department of Physics and Astrophysics, 
University of Delhi, New Delhi 110007, India}

\begin{abstract}
%
%{}\\
The mass of an astrophysical object can be estimated by the amount 
of gravitational lensing of another object that it causes. To arrive 
at the estimation however, one assumes the validity of the inverse 
square law of gravity, or equivalently an attractive $1/r$ potential.
We show that the above, augmented by a logarithmic potential at 
galactic length scales, proposed earlier to explain the flat galaxy 
rotation curves, predicts a larger deflection angle for a given 
mass. In other words, the true mass of the object is less than its 
estimated value. This may diminish the importance and role of dark 
matter in explaining various observations. 

\end{abstract}
%\pacs{123xxx}

\maketitle

The theory of gravitation -- Newtonian or relativistic, is one of the most 
successful physical theories to be formulated. It is often assumed that 
General Relativity (GR) is valid as such at all length scales, from the 
planetary to the galactic and cosmological. But is this really true? While 
Newton's gravity and GR have been tested to a very high degree of accuracy 
at planetary scales, e.g. in the verification of Kepler's laws and the 
perihelion precession of Mercury
\cite{tests1,tests3},
things become less clear as one goes further out. For instance, at galactic 
scales, one has to introduce dark matter, 
%in addition 
to add to the luminous matter and make the standard gravity theory `work', 
% \cite{gm1,gm2,gm3,gm4}, 
while at cosmological length scales, one needs to bring in dark energy 
for the same reason 
\cite{weinberg}. 
With these new and speculative elements in place, GR stays intact, but 
the price one pays is that the origin of dark matter and dark energy 
remain unknown. 

%\vspace{-0.15cm}
One can take an alternative point of view that the law of gravitation should 
be such that all astrophysical and cosmological observations are explained 
without the need for introducing unobserved elements such as dark matter 
and dark energy. This is possible, if one postulates the gravitational 
field and potential 
%$V(r)$ 
on a test particle of unit mass 
%$m$ 
to be as follows:
\begin{eqnarray}
E (r) &=& - \fr{k_1}{r^2} - \frac{k_2}{r} + k_3 r \,,
\label{field1} \\
V(r) &=& - \fr{k_1} r \,+\, k_2 \,\ln \le(\fr r {r_0}\ri)
\,-\, \fr 1 2 \,k_3\, r^2 \,;
%\quad 
%~~0\leq r < \infty \,,~~ 
\label{pot4}
%\label{pot1}
\end{eqnarray}
with $0\leq r < \infty$, and where 
%
%$r_1$ and $r_2$ are respectively the galactic and the cosmological length scales, 
%
$k_1$, $k_2$ and $k_3$ are positive-valued constants, and $r_0$ is an 
arbitrary length scale, which does not enter in the force law 
and hence has no 
observable consequence. 
We may write $k_1 = G M$ and $k_2 = \l M$, where
$G$ denotes the Newton's constant, $M$ the gravitating mass and $\l$
another dimensionful and positive-valued constant
\footnote{Note that the $r^2$ term in Eq.(\ref{pot4}) and Eq.(\ref{pot6}) below can be 
%identified with 
treated as the cosmological constant ($\L$) term in Newtonian cosmology, 
%for which the following identification can be made
under the identification $k_3=\Lambda\,c^2/3$.}.

Note that Eqs.(\ref{field1}) and (\ref{pot4}) are simply {\it proposals} 
for the gravitational field and potential stretching across all length 
scales and their correctness will ultimately depend on their predictions 
matching with current or future observations. 
It may be emphasized that the above additions to the Newtonian terms are 
not `effective' or a consequence of a given matter distribution. These 
are postulated to be truly fundamental, at the same level as the attractive 
inverse-square law. 
%%%%%%%%%%%%%%%%%%%%%%
The proposal can be motivated from other considerations as well. 
For example, if the gravitational 
%constant 
coupling factor $G$ has a small spatial variation, then writing it 
as 
%$G+G' r + G'' r^2/2 + \dots$, 
%\tb{
%
\be \label{G-expand}
G(r) =\, G(0) +\, r G'(0) +\, r^2 G''(0)/2 +\, \dots \,,
\ee 
where the primes denote the $r$-derivatives,
%evaluated at $r=0$, 
%}
one obtains the series in Eqs.\,(\ref{field1}) and (\ref{pot4})
(the careful reader will notice that the $G''$ term 
%is actually omitted in 
%\tb{
may actually be omitted from
%}
%
the above, as the resultant $r$-independent force does not have any 
clear physical meaning.).
Furthermore, it can be easily seen that the variation in $G$ 
required to obtain Eqs.\,(\ref{field1}) and (\ref{pot4}) with the 
coefficients of the right magnitudes is consistent with 
observational bounds on the same 
\cite{gprime1,gprime2,gprime3}.
%%%%%%%%%%%%%%%%%%%%%%
%A complete (Newtonian or relativistic) theory for this potential will be studied in a future publication. 
%
%we write $k_1= G M$ and $k_2 = \l M $, $k_3 $ is positive and $r_0$ is an arbitrary length scale, which does not enter in the force and hence has no observable consequence.
%
%%%%%%%%%%%%%%%%%%%
%Of course, one has $k_1=GMm$, where $M$ is the lens mass.
%%%%%%%%%%%%%%%%%%%%
%
%Writing $k_1=GM$ and $k_2=\lambda\,M$, where
%$M$ is the lens mass and $\lambda$ a constant
%
%

%Such a potential has been studied by various authors, e.g. in \cite{binney1,binney2,bobylev,sivaram,haranas}, and the orbits in this potential have been computed in \cite{valluri}.

%\tb{
While we plan to look for a complete (Newtonian or relativistic) 
theory for the potential $V(r)$ given by Eq.\,(\ref{pot4}) in a
future work, let us mention here that such a potential has been 
studied by various authors, e.g. in
\cite{binney1,binney2,bobylev,sivaram,haranas},
and the orbits in this potential have been computed in 
\cite{valluri}.
%}
%
It is clear from the above functional dependence of $V(r)$ that
%It is clear from the nature of the above function that
there exist regions defined by the length scales $r_1$ and $r_2$,
%such that each term in Eq.(\ref{pot4}) dominates in a certain region, 
such that each term in Eq.(\ref{pot4}) dominates in a certain region, which can 
therefore be re-written as
%
%approximately as 
%and Eq.(\ref{pot4}) reduces to approximately 
%
%\bea
%V(r) &=& V_1 = - \fr{k_1} r \,, \quad r\leq r_1 \,, \label{pot1} \\
%%
%&=& V_2 = k_2 \,\ln \le(\fr r {r_0}\ri) , \quad r_1 < r\leq r_2 \,, 
%\label{pot2} \\
%%
%&=&  V_3 = - \fr 1 2 \,k_3\, r^2 \,, 
%%\frac{1}{2}\,m\, \omega^2 r^2 ,
%\quad r_2 < r < \infty \,. \label{pot3}
%%
%\eea
%
%
%
\bea 
V (r) \,=\, \le\{ \begin{aligned}
&V_1 (r) = - \dfrac{k_1} r \,; ~ &r\leq r_1 \,,
%\label{pot1} 
\\
&V_2 (r) = k_2 \,\ln \le(\dfrac r {r_0}\ri) ; ~ &r_1 < r\leq r_2 \,, 
\label{pot6} 
\\
&V_3 (r) = - \dfrac 1 2 \,k_3\, r^2 \,; ~
%\frac{1}{2}\,m\, \omega^2 r^2 ,
& r_2 < r < \infty \,. 
%\label{pot3}
\end{aligned} 
\ri. 
\eea 
%
%It is clear that 
The potential $V_1$, 
%of course, 
or the corresponding force $\,|F_1|=|\vec\nabla\,V_1|=k_1/r^2\,$,
explains all gravitational phenomena at planetary scales ($r\leq r_1$), 
while $r_1$ and $r_2$ signify the galactic and cosmological scales, 
respectively.
%namely 
Specifically, $\, r_1 \simeq 10^{21}\,$m $=30$ kpc and 
$\, r_2 \simeq 10^{26}\,$m $=3\times 10^6$ kpc.
The potential $V_2$, on the other hand, in the intermediate distance
range $r_1 \leq r_2$, gives rise to flat rotation curves of galaxies.
This is because the corresponding force $\,|F_2|=|\vec\nabla\,V_2|
= k_2/r\,$ has the same $r$ dependence as the centripetal force 
$m v^2/r$, and therefore when they are equated, it results in a 
constant speed $v$ (of a test particle of mass $m$).  
Furthermore, using $|F_1|=|F_2|$ at $r\simeq r_1$, since this is 
where $V_1$ gives way to $V_2$, one gets 

\vspace{-5pt}
\be \label{lambda}
\l =\, \fr G {r_1} \,. 
\ee 
Finally, for suitable values of the constant $k_3$, the potential 
$V_3$ can lead to an accelerated expansion of our universe.

%
%%%%%%%%%%%%%%%%%%%%%%%% figure 1 %%%%%%%%%%%%%%%%%%%%%%%%
\begin{figure}[!htp]
\centering
\includegraphics[height=2.5in,width=3.37in]{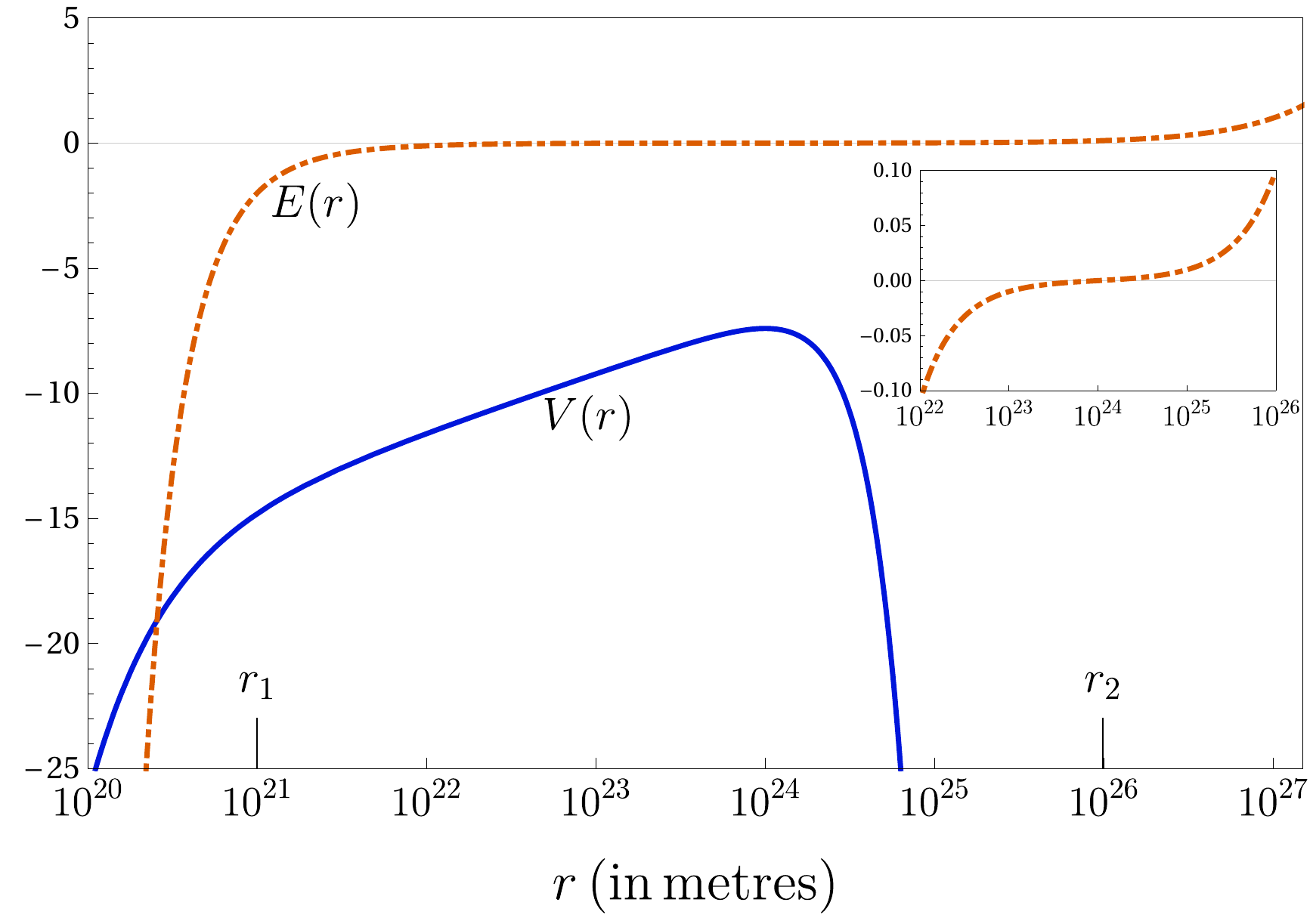}
\caption{\footnotesize Functional profiles of the total potential 
$V(r)$ and the corresponding force field $E(r)$, for $r$ in the 
log$_{10}$ scale encompassing the range $10^{20}\,$--$\,10^{27}\,$m. 
The long ticks at $10^{21}\,$m and $10^{26}\,$m mark the typical 
scales $r_1$ and $r_2$ respectively. The plots are generated by 
making a further scale setting $r_0 = 10 r_2$, with an exaggerated 
depiction of $E(r)$ for $r \in[10^{22},10^{26}]\,$m in the inset.}
\label{LP_Fig}
\end{figure}
%%%%%%%%%%%%%%%%%%% figure 1 ends %%%%%%%%%%%%%%%%%%%%%%%%%
%
%\tb{
Fig.\,\ref{LP_Fig} shows a typical profile of $V(r)$ and 
correspondingly that of the function $E(r)$, for $r$ in the 
range $10^{20}\,$--$\,10^{27}\,$m. For the purpose of 
illustration, the constants and scales are chosen as: 
(i) $k_1 = 1$ in units of $GM$, 
(ii) $k_2 = k_1/r_1$, where $r_1 = 10^{21}\,$m, and 
(iii) $k_3 = k_1/(r_0 r_1^2)$, with the stipulation 
$r_0 = 10 r_2$, where $r_2 = 10^{26}\,$m. 
Such a stipulation ensures that the force due to the 
logarithmic potential $V_2$ remains attractive even beyond 
the length scale $r_2 = 10^{26}\,$m, to a fair extent. The 
overall force field $E(r)$ though, changes from being 
attractive to repulsive around $r \simeq 10^{24}\,$m, as 
noticed from the inset of Fig.\,\ref{LP_Fig}. Nevertheless, 
the repulsive nature of the force, due to the dominance of 
the potential $V_3$, starts to become more and more 
prominent only when $r \to r_2 = 10^{26}\,$m. 
%
%Note also that the $r$-axes in Fig.\,\ref{LP_Fig} and its inset are shown in the log$_{10}$ scale.
%}

In this work, however, we limit ourselves to 
%we will be concerned with only 
the potential
%s $V_1$ and $V_2$, i.e. 
%we use at 
$V(r) \simeq V_1(r) + V_2(r)$ in the range $0 \leq r \leq r_2$.
%\tb{
Whether this potential can lead to stable closed orbits of 
test particles 
%(say, the massive ones) 
is what one may ask in the first place. This amounts to examining 
the existence of minimum point(s) of the effective potential 
$V_{\text{\scriptsize eff}}(r) = V(r) + \ell^2/(2 r^2)$, where 
$\ell$ denotes the particle's angular momentum per unit mass. 
If there exists a minimum at say, $r = r_c$, then that implies 
the existence of a stable circular orbit of radius $r_c$. Now, 
given the forms of $V_1$ and $V_2$ in Eq.\,(\ref{pot6}), it is 
easy to check that the existence of a minimum of 
$V_{\text{\scriptsize eff}}$ requires the following conditions
to be satisfied
\be \label{stab-cond}
k_2 r_c^2 +\, k_1 r_c =\, \ell^2 \,, \qquad k_1 r_c < 2 \ell^2 \,.
\ee 
While the first condition demands the only plausible radius to
be
\be \label{stab-rad}
r_c =\, \fr{k_1}{2 k_2} \le[\sq{1 +\, \fr{4 \ell^2 k_2}{k_1^2}}
-\, 1\ri] \,,
\ee 
the second condition implies $4 \ell^2 k_2/k_1^2 > 0$, or simply
$k_2 > 0$. Therefore, circular orbits are {\em always} stable,
for any given pair of values of the constants $k_1$ and $k_2$,
with the latter presumed to be positive-valued. 
\footnote{This is similar to the purely Newtonian case (or in
general when the total force varies as $r^{-n}$ with $n < 3$). 
However, all circular orbits are not stable when the force is 
effectively Newtonian plus a term $\sim r^{-s}$ with $s > 3$, 
for e.g. in Schwarzschild space-time.}
%The requirement can in general be stated as the following conditions on the corresponding force $\,F(r) = - dV/dr\,$:
%$\,F(r) = - k_1/r^2 - k_2/r\,$ 
%
%\be \label{F-cond}
%F(r_c) =\, - \fr{m \ell^2}{r_c^3} \,, \qquad 
%\fr{dF}{dr}\Big\vert_{r_c} \!+ \fr 3 {r_c} F(r_c) <\, 0 \,.
%\ee 
%
%It is obvious that a force term given as an inverse radial power ($1/r^n$) would lead to stable circular orbits as long as the power $n < 3$. If we consider the logarithmic potential $V_2$ alone, then that corresponds to $n = 1$, and hence the stable orbits are guaranteed. However, for the force due to the total potential $V = V_1 + V_2$, one requires to carry out a rather careful check. 
%}

%\tb{
Let us now turn our attention to the unbound orbits and
%} 
compute the deflection angle of light due to a gravitational 
lens of mass $M$,
%(i.e. approximated by a point mass), 
for e.g. following
\cite{carroll,cheng}. 
Note that one does not need to assume the validity of the 
Poisson equation in the derivation. One just requires the 
form of the potentials and use the fact that light rays follow 
geodesics. 
The standard deflection angle
\footnote{We assume here that the photon and other test particles follow geodesics and that any modification of the effective Newtonian potential can be incorporated in the metric components, at least to the leading order,
as in General Relativity.}
\bea \label{delta1}
\d_1 &=& \fr 2 {c^2} \int \nabtr V_1 \, ds \nn \\ 
&=& \fr{2 G M b}{c^2} \int_{-\infty}^{\infty} \fr{dx}
{(b^2+x^2)^{3/2}} \nn \\
&=& \fr{4GM}{b c^2} \,,
\eea 
with $b$ as the impact parameter, is now replaced by
%
%\begin{eqnarray}
%\delta_1 &&=\frac{2}{c^2} 
%\int \vec\nabla_{\perp} V_1 \, ds \\ 
%
%&& = \frac{2GMb}{c^2}\,
%\int_{-\infty}^{\infty} \frac{dx}{(b^2+x^2)^{3/2}} \nonumber  \\
%
%&&= \frac{4GM}{bc^2}
%\end{eqnarray} 
%
%is now replaced by 
%
\bea \label{delta4} 
\d &=& \fr 2 {c^2} \int \nabtr \le(V_1 + V_2\ri) ds \nn \\ 
&=& \d_1 + \fr{2\l\,M b}{c^2} \int_{-\infty}^\infty 
\fr{dx}{(b^2+x^2)} \nn \\
&=& \fr{4GM}{b c^2} +\, \fr{2 \l \, M \pi}{c^2} \nn \\
&\equiv& \fr{4GM'}{b c^2} \,, 
\eea

\vspace{-5pt}
\noindent
where

\vspace{-15pt}
\be 
M' \equiv \, M \left( 1 + \epsilon\right) \,, \quad 
\mbox{with} \quad \epsilon = \fr{\l \, b \pi}{2 G} \,.
\label{mass1}
\ee
%
%That is,  
%
%\be \label{delta4a}
%\d =\, \fr{4GM'}{b c^2} \,,
%\ee
%where 
%% 
%\be \label{mass1}
%M' \equiv\, M + \D \,, \quad \mbox{with} \quad 
%\D = \fr{\l \, M b \pi}{2 G} \,. 
%\ee 
%%
%
%\vspace{-0.9cm}
%\be \label{mass1}
%\text{where~}~~
%M' \equiv\, M \left( 1 + \epsilon\right) %\,, \quad \mbox{with} \quad 
%\epsilon = \fr{\l \, b \pi}{2 G} \,. 
%\ee 
%%
%
%The total deviation thus consists of the usual GR part ($\d_1$) 
%and an additional piece 
%
%\be \label{delta2}
%\d_2 \equiv\, \d - \d_1 =\, \fr{2 \l \, M \pi}{c^2} \,,
%\ee 
%
%due to the logarithmic term in Eq.(\ref{pot4}).
%
It can be seen from the above that whereas $M$ is the actual 
gravitating mass, $M'$ is the mass estimated via lensing, if 
one ignores the logarithmic term.
%if the logarithmic potential term is ignored. 
%Eq.(\ref{delta4}) can be inverted to give
%
%\begin{eqnarray}
%\delta &&= \delta_1 + \delta_2 \\
%
%\delta_1 
%&&= \frac{2}{c^2} 
%\int \vec\nabla_{\perp} 
%\left( V_1 + V_2 \right) \, ds \\ 
%
%&&= \frac{2GMb}{c^2}\,
%\int_{-\infty}^\infty \frac{dx}{(b^2+x^2)^{3/2}} 
%+ %\frac{2k_2 b}{c^2}\,
%\frac{2\lambda\,M\,b}{c^2}\,
%\int_{-\infty}^\infty \frac{dx}{(b^2+x^2)}  \nonumber \\
%
%&&= \frac{4GM}{b c^2} + %\frac{2k_2\pi }{c^2}
%\frac{2\,\lambda\,M\,\pi}{c^2}
%~.\label{delta4} \\
%
%&&\equiv \frac{4GM'}{b c^2}~,
%\end{eqnarray}
%
%where,
%
%\begin{eqnarray}
%\delta_2 = 
%\frac{2k_2\,\pi  }{c^2}
%\frac{2\,\lambda\,M\,\pi  }{c^2}~~.
%\end{eqnarray}
%
%It can be seen that the total deviation consists of the usual Newtonian 
%part ($\delta_1$) and an additional piece ($\delta_2$), due to the 
%logarithmic term. In the above, we have defined 
%
%\begin{eqnarray}
%M' &&= M\,\left[ 1 + 
%\frac{k_2\, b\,\pi}{2GM}
%\frac{\lambda\, b\,\pi}{2G}
%\right] \label{mass1} \\
%
%&&\equiv  M +  \Delta \\
%
%\text{where,}~~\Delta &&= %\frac{k_2\, b\, \pi }{2G}
%\frac{\lambda\,M\, b\, \pi }{2G}~.
%\end{eqnarray}
%
%Where $M$ is the true mass while $M'$ is the mass estimated via lensing, 
%if one ignores the $k_2$ or the logarithmic potential term. 
%Eq.(\ref{delta4}) can be inverted to give
%
%\begin{eqnarray}
%M = 
%\frac{bc^2}{4G}\,
%\left[ 1 + \frac{\pi \lambda b}{2G} \right]^{-1}
%\left( \delta - %\frac{2k_2\,\pi}{c^2} 
%\frac{2\,\lambda\,M\,\pi}{c^2} 
%\right) 
%\label{mass2}
%\end{eqnarray}
%
%
Consequently, it follows from Eqs.(\ref{delta4}) and (\ref{mass1}), 
%and (\ref{mass2}), 
that ignoring the logarithmic potential term results in the following:
\ben[(i)]
\item for a given mass, the deflection angle would be underestimated, 
\item for a given deflection angle, the computed mass would be 
overestimated. 
\een 
%
%\vspace{-0.2cm}
Similarly, the magnification of the lens, which proportional to the Einstein radius (for near alignment of source, lens and observer), now reads \cite{cheng}
\begin{eqnarray}
\theta_E &&= \sqrt{\frac{D_{LS}}{D_L D_S}\,\frac{4GM'}{c^2}}~ \\
&&= \sqrt{\frac{D_{LS}}{D_L D_S}\,
\frac{4GM(1+\epsilon)}{c^2}}~ 
\end{eqnarray}
where $D_S,D_L$ and $D_{LS}$ are respectively the distances from the observer to the source, lens and the distance from the source to the lens respectively. Again, this is more than what would be the case without the logarithmic term, and conversely, the true mass would be less than that estimated without the term.  
%%

%\vspace{-0.2cm}
Using Eq.(\ref{lambda}), we have the error in the mass estimation 
given by
%The error in mass estimation is given by

\vspace{-15pt}
\be
\e
%= \frac{\Delta}{M}
%\simeq \frac{M_2}{M_1} 
= \fr{\l \, b \pi}{2G} = \fr{b \pi}{2r_1} \,.
\ee
%
%where we have used Eq.(\ref{lambda}). 
This 
%simple formula 
shows that the discrepancy increases with distance of closest approach
$b$, and 
%as we shall see below, 
becomes significant at galactic length scales.

%\vspace{-0.15cm}
Let us now 
%consider 
%\tb{
look into
%}
some specific cases.
%
%For
%\tb{
Consider first the
%}
deflection of light by the Sun at a grazing incidence.
%, we plug-in
%\tb{
For this, plugging in
%}
the well-known value $\, b = 2.3\times 10^{-11}$ kpc, 
%to get
%\tb{
we get
%}
%
\be
\e =\, 1.2 \times 10^{-12} \,,
\ee
which, although non-zero, is likely beyond the scope of current 
measurements 
\cite{accuracy}.
In this case, a mass estimation via lensing using only the Newtonian 
term would be quite accurate. 

%\vspace{-0.15cm}
%Next, we consider 
%\tb{
Consider next
%}
a galaxy or a cluster of galaxies with $b \simeq 10$ to $100$ kpc, 
of the order of the galaxy (or cluster) size.
%$b \approx 10^{21}$ m, 
In this case, we get
%
%\vspace{-5pt}
%
\be
\e \simeq\, 1~\text{to}~10 \,. 
\ee
In other words, estimating galaxy masses using gravitational 
lensing without taking into account the logarithmic term will 
give an error ${\cal O}(1) - {\cal O}(10)$
\footnote{
Modified Newtonian Dynamics (MOND) also effectively gives 
rise to a logarithmic potential, but with $k_2 = \sqrt{GM a_0}$
\cite{MOND1,MOND2,MOND3,MOND4,MOND5}, 
where $a_0 = 1.2\times 10^{-10}~m/s^2$. This gives rise to 
$\e = {\pi b \sq{a_0}}/{\sq{GM}}$, which for 
$M = \le(10^{12} - 10^{14}\ri) M_\odot$ [where $M_\odot$ is 
the solar mass] also translates to $\e = 1$ to $10$. This 
explains why MOND seems to work (only) at galactic scales. 
}.
This error is applicable, for example, to the estimation of 
the bullet cluster of mass in the above range 
\cite{bullet1,bullet2}. 
In particular, since lensing from this galaxy cluster has been used to 
infer the presence of dark matter of mass comparable to the above, it 
seems quite likely that such a mass has been over-estimated. In fact, 
this shows that the potential $V_2$, introduced to explain the flat 
galaxy rotation curves, may also get rid of any extra, non-luminous 
matter in the galaxy cluster, thereby providing a reassuring 
consistency check of the model. This and the apparent lack of dark 
matter in certain galaxies
\cite{nodm1,nodm1a,nodm2,nodm3,nodm4,nodm5}
shows that at the very least, an analysis of available data in light 
of this new potential ought to be examined. 

It is natural to speculate on the origin of the terms in addition to 
the Newtonian term in the potential given in Eq.(\ref{pot4}). While 
it is possible that they are emergent from a more fundamental origin 
\cite{qp1,qp2,qp3,qp4,qp5,qp6}, 
at this stage, as stated in the introduction, we simply consider 
their possibility beyond the scales at which the Newtonian potential 
has been measured directly, and examine its implications. A covariant 
version of this theory is in order, which we will examine in due 
course in future. 
%
%\tb{
Further implications of our model for a host of other astrophysical 
and cosmological scenarios, as well as statistical analyses of observed 
phenomenon in the light if the new proposal should be studied, which we 
hope to do in future works as well.
%}

In summary, our study suggests that it is possible after all, that 
dark matter may play a less important role in explaining certain 
astrophysical phenomena, than originally thought. 
%
%\tb{
Of course, there remains a need to address certain related issues,
for instance, the well-known offset in the luminous and dark matter 
mass distributions in the bullet cluster. While attempts have been
made to explain this in some variants of the MOND theory (see for e.g.
\cite{MOND3,MOND4,MOND5}
and references therein), and also by introducing the concept of 
{\it gravitational retardation} in refs.
\cite{yahalom1,yahalom2,yahalom3},
we need to find a way do so in order to have a reconciliation of 
the proposal we have put forward in this paper. Note however that 
the approach we have taken is motivated by a spatial variation of 
the Newtonian coupling factor $G$, with specified length scales 
for the dominance of the different terms of the total effective 
potential.
This is in sharp contrast with MOND, which introduces a scale 
pertaining to the acceleration. Gravitational retardation, on 
the other hand, is a natural requirement which can be reckoned
within the purview of our model. We do hope to align our model 
to refs.
\cite{yahalom1,yahalom2,yahalom3},
with a concrete methodology in mind, which is to have the scale 
$r_1$ (when the logarithmic potential $V_2$ becomes significant) 
comparable to the retardation scale introduced therein. 
It would nevertheless be taken up as a potential future project, 
involving a significant amount of work, as part of which we have 
to estimate the additional retardation effect arising from the 
potential term $V_2$, as well as propose a time-dependent 
density distribution profile to match up with the observational 
results.
%}
%
%that one may 
%not need to introduce dark matter to explain the observations of our Universe. 
%

%
%Eq.(\ref{delta4}) can be inverted to give
%
%\begin{eqnarray}
%M &&= \frac{b c^2}{4G} \left( \delta_0 - \frac{2k_2\pi}{c^2} \right) ~.
%\label{delta5} \\
%%%
%&&\equiv M_1 - M_2~.
%\end{eqnarray}
%
%The error in such overestimation is 
%
%\begin{eqnarray}
%\epsilon = \frac{M_2}{M_1} = \frac{2 k_2\pi}{\delta_0 c^2}~.
%\end{eqnarray}
%%
%\begin{eqnarray}
%\epsilon = \frac{2\pi\,\sqrt{GMa_0}}{\delta_0 \,c^2}~,
%\end{eqnarray}
%%
%

%\vspace{5pt}
%\noindent 
%{\bf Acknowledgment}
\section*{Acknowledgment}
\no 
We thank S. J. Landau for useful comments. We thank the anonymous referees for their useful comments which have helped in improving the paper. 
This work is supported by the Natural Sciences and Engineering
Research Council of Canada. SS acknowledges financial support 
from Faculty Research Programme Grant -- IoE, University of 
Delhi (Ref. No./IoE/2021/12/FRP).
%

%%%%%%%%%%%%%%%%%%%%%%%%%%%%%%%%%%%%%%%%%%%%%%%%%%%%%

\clearpage

%\section*{References}

\end{document}